\documentclass[allclo]{FBSart}
\usepackage{amsfonts}
\usepackage{amssymb}
\usepackage{epsfig}

\title{Electromagnetic properties of diquarks}
\author{P. Maris\thanks{\textit{E-mail address:} 
pim6@pitt.edu}}
\institute{Dept. of Physics and Astronomy, 
University of Pittsburgh, Pittsburgh, PA 15260}

\runningauthor{P.\,Maris}
\runningtitle{Electromagnetic properties of diquarks}
\begin{document}

\maketitle
\begin{abstract}
Diquark correlations play an important role in hadron physics.  The
properties of diquarks can be obtained from the corresponding bound
state equation.  Using a model for the effective quark-quark
interaction that has proved successful in the light meson sector, we
solve the scalar diquark Bethe--Salpeter equations and use the
obtained Bethe--Salpeter amplitudes to compute the diquarks'
electromagnetic form factors.  The scalar $ud$ diquark charge radius
is about 8\% larger than the pion charge radius, indicating that these
diquarks are somewhat larger in size than the corresponding mesons.
We also provide analytic fits for the form factor over a moderate
range in $Q^2$, which may be useful, for example, in building
quark-diquark models of nucleons.
\end{abstract}

\section{\label{sec:intro}
Diquarks}
Diquarks are coloured states of two quarks.  Due to their colour
charge, they are expected to be confined, and cannot be observed
directly.  Despite being confined, they can play a role in the
dynamics of baryons: two quarks in a colour anti-triplet configuration
can couple with a third quark to form a colour singlet baryon.  Indeed
quark-diquark models have been used quite successfully in various
models for baryons~\cite{bentz,tuebingen,argonne}.  More recently,
they also have been suggested as the constituents of pentaquarks: one
could imagine a pentaquark as a bound state of two diquarks and an
antiquark~\cite{Jaffe:2003sg}, or as a bound state of a diquark and
a triquark~\cite{Karliner:2003dt}.

For calculations of the baryon spectrum within a quark-diquark model,
the most important property of the diquarks is their mass.  There are
a number of different calculations of these masses available, both
from lattice QCD~\cite{Hess:1998sd} and using the set of
Dyson--Schwinger equations
[DSEs]~\cite{Cahill:1987qr,Burden:1996nh,Maris:2002yu}.  These
calculations seem to agree with each other to within about 20\%.  For
calculations of electromagnetic properties of nucleons (and other
baryons), one also needs to know how photons couple to diquarks.
Since they are not observable, we cannot obtain this information from
experiments (at least not directly), and we have to rely on
theoretical calculations of the electromagnetic properties of
diquark correlations.

Here we discuss the electromagnetic form factors of the scalar
diquarks, using the DSE approach~\cite{Roberts:1994dr,Roberts:2000aa,
Alkofer:2000wg,Maris:2003vk}.  The model we use for the interaction
has been shown to give an efficient description of the light
pseudoscalar and vector mesons~\cite{Maris:1997tm,Maris:1999nt}.  In
particular the pion and kaon electromagnetic form factors are in
excellent agreement with the available experimental
data~\cite{Maris:emf}.  Our calculation can serve as a guide to
further improvements of quark-diquark models of nucleons and other
hadrons.

\section{\label{sec:DSEapproach}
Dyson--Schwinger Approach}
The DSE for the renormalized quark propagator in Euclidean space is
\begin{eqnarray}
  S(p)^{-1} &=& i \, Z_2\, /\!\!\!p + Z_4\,m_q(\mu) 
    +  Z_1  \int_q^\Lambda\!  g^2 D_{\mu\nu}(k) \, 
        {\textstyle\frac{\lambda^i}{2}} \gamma_\mu \, 
    S(q) \, \Gamma^i_\nu(q,p) \; ,
\label{gendse}
\end{eqnarray}
where $D_{\mu\nu}(k=p-q)$ is the dressed-gluon propagator,
$\Gamma^i_\nu(q,p)$ the dressed-quark-gluon vertex with colour-octet
index $i=1,\ldots,8$, and $Z_2$ and $Z_4$ are the
quark wave-function and mass renormalization constants.  The notation
\mbox{$\int^\Lambda_q \equiv \int^\Lambda d^4 q/(2\pi)^4$} stands for
a translationally invariant regularization of the integral, with
$\Lambda$ being the regularization mass-scale.  The regularization is
removed at the end of all calculations, by taking the limit
\mbox{$\Lambda \to \infty$}.  We use the Euclidean metric where
\mbox{$\{\gamma_\mu,\gamma_\nu\} = 2\delta_{\mu\nu}$},
\mbox{$\gamma_\mu^\dagger = \gamma_\mu$} and \mbox{$a\cdot b =
\sum_{i=1}^4 a_i b_i$}.  The most general solution of
Eq.~(\ref{gendse}) has the form \mbox{$S(p)^{-1} = i /\!\!\! p A(p^2)
+$} \mbox{$B(p^2)$} and is renormalized at spacelike $\mu^2$ according
to \mbox{$A(\mu^2)=1$} and \mbox{$B(\mu^2)=m_q(\mu)$} with $m_q(\mu)$
being the running current-quark mass.

For practical calculation, we have to make a truncation of the set of
DSEs~\cite{Roberts:1994dr,Roberts:2000aa,Alkofer:2000wg,Maris:2003vk}.
Here we adopt the rainbow truncation of the quark DSE, given by
\begin{equation}
\label{ourDSEansatz}
Z_1 \, g^2 D_{\mu \nu}(k) \Gamma^i_\nu(q,p) \rightarrow
 {\cal G}(k^2) D_{\mu\nu}^{\rm free}(k)\, \gamma_\nu
                                        \textstyle\frac{\lambda^i}{2} \,,
\end{equation}
where $D_{\mu\nu}^{\rm free}(k=p-q)$ is the free gluon propagator in
Landau gauge and ${\cal G}(k^2)$ a phenomenological effective
interaction~\cite{Maris:1997tm,Maris:1999nt}.

\subsection{Diquarks}
Diquarks are coloured quark-quark correlations.  Two quarks can be
coupled in either a colour sextet or a colour anti-triplet.  Single
gluon exchange leads to an (effective) interaction that is attractive
for diquarks in a colour anti-triplet configuration, but the
interaction is repulsive in the colour sextet
channel~\cite{Cahill:1987qr}.  Furthermore, it is the diquark in a
colour anti-triplet that can couple with a quark to form a
colour-singlet baryon.  Thus we will only consider the diquark
correlations in a colour anti-triplet configuration.

Using the antisymmetric tensor $\epsilon_{abc}$ as a representation of
the anti-triplet, the diquarks are described by solutions of the
homogeneous Bethe--Salpeter equation [BSE]
\begin{eqnarray}
 \Big(\Gamma_D(p_+,-p_-)C\Big)^{\alpha\beta'} \epsilon_{abc} &=& 
    -\epsilon_{dec} \int_q^\Lambda\! 
    K_{ad;be}^{\alpha\gamma;\beta\delta}(p_+,q_+;-p_-,-q_-) 
    \, C^{\delta''\delta} \, C^{\beta\beta'}
\nonumber \\ && 
    S^{\gamma\gamma'}(q_+) \, 
    \Big(\Gamma_D(q_+,-q_-)C\Big)^{\gamma'\delta'} \, 
    S(q_-)^{\delta'\delta''} \, .
\label{eq:DiqBSE}
\end{eqnarray}
where $p_+ = p + P/2$ and $-p_- = -p + P/2$ are both outgoing quark
momenta (and similarly for $q_+$ and $-q_-$), and $C$ is the charge
conjugation matrix $C$, which satisfies $C^2 = -1$ and $
C\,\gamma_\mu\,C \,=\, (\gamma_\mu)^T$.

In ladder truncation this equation reduces to
\begin{eqnarray}
 \Gamma_D(p_+,-p_-) \, C &=&  -\frac{2}{3}\,
        \int_q^\Lambda\! {\cal G}(k^2) \; D_{\mu\nu}^{\rm free}(k)
    \gamma_\mu \, S(q_+)  \, \Big(\Gamma_D(q_+,-q_-) \, C\Big) \, 
    S(q_-) \, \gamma_\nu \; .
\nonumber \\ &&
\label{eq:ladderBSEd}
\end{eqnarray}
This equation resembles the bound state BSE for a meson, the only
difference being that the effective interaction in the diquark BSE is
reduced by a factor of two compared to that in the meson BSE in ladder
truncation.

The allowed flavour structure of the diquarks follows from the Pauli
principle.  The total wave function has to be antisymmetric in all its
labels, and we know already that they are antisymmetric in their
colour indices.  For three flavours that implies that the scalar
diquarks have to be in a flavour anti-triplet.  On the other hand,
axial-vector diquarks are in a flavour sextet.  Convenient flavour
matrices for the anti-triplet states are the Gell--Man matrices
$\lambda_2$, $-\lambda_5$, and $\lambda_7$.  Explicitly, that means
that we consider only $ud$, $us$, and $ds$ scalar diquarks here.

Different types of diquarks are characterized by different Dirac
structures.  Since the intrinsic parity of a quark-quark pair is
opposite to that of a quark-antiquark pair, the scalar diquark
Bethe--Salpeter amplitude [BSA], or rather
$\Gamma_{D,{\hbox{\scriptsize{scal}}}}(q_+,-q_-)C$, has the general
decomposition
\begin{eqnarray}
\label{diquarkdecomp}
\Gamma_{D,{\hbox{\scriptsize{scal}}}}(q_+,-q_-)C 
&=& \gamma_5 \big[ i E(q^2,q\cdot P) 
        + \;/\!\!\!\! P \, F(q^2,q\cdot P) 
\nonumber \\ && {}
        + \,/\!\!\!k \, G(q^2,q\cdot P) 
        + \sigma_{\mu\nu}\,P_\mu q_\nu \,H(q^2,q\cdot P) \big]\,,
\end{eqnarray}
which is the same form as a pseudoscalar meson BSA.

\subsection{Electromagnetic interactions}
\begin{figure}[tbh]
\begin{center}
\epsfig{file=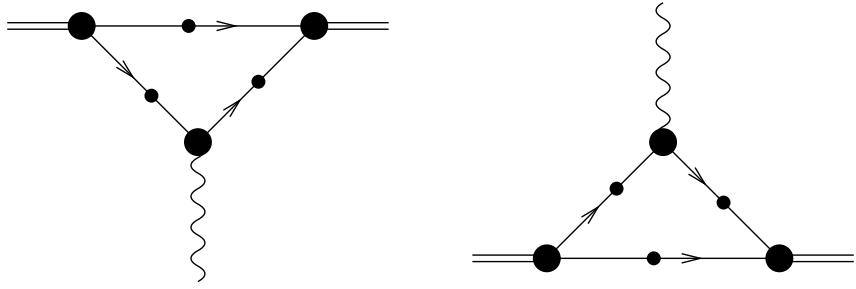,height=120mm,angle=-90}
\end{center}
\caption[]{\label{fig:triangle}
Triangle diagrams representing the diquark-photon 
interaction in impulse approximation.}
\end{figure}

The coupling between a photon and a diquark with flavour structure $ab$
is in impulse approximation described by two very similar diagrams,
each with the photon coupled to one of the two quarks, depicted in
Fig.~\ref{fig:triangle}.  
The diagram where the photon couples to quark $a$ of the scalar
diquark is given by the loop integral
\begin{eqnarray}
  I_\mu^{ab;a}(P,Q) &=& \int\!\!\frac{d^4k}{(2\pi)^4} \,
    {\rm Tr}\bigg[ S_a(k_{++}) \Gamma_\mu(k_{++},k_{+-}) S_a(k_{+-}) 
\nonumber\\ && \times 
      \Gamma_{D,ab}(k_{+-},-k_{-})C 
      S_b(k_-) C\bar{\Gamma}_{D,ba}(-k_{-},k_{++}) \bigg]
\end{eqnarray}
where $Q$ is the photon momentum, and the incoming and outgoing
diquarks have momentum $P \mp Q/2$; the internal momenta are $k_- =
k-P/2 $, and $k_{+\pm} = k + P/2 - Q/2$; the (repeated) flavour index
after the semicolon in $I_\mu^{ab;a}(P,Q)$ indicates which of the
quarks is coupled to the photon.  The expression for the other diagram
is very similar.  These two contributions have to be added together
with the appropriate quark charge.  Notice the difference with the
expression for the meson-photon coupling: there is no factor $N_c$
multiplying the loop integral.  However, this factor $N_c$ is also
absent from the normalisation condition for the diquarks (when
compared to that of meson bound states), and thus effectively this
difference cancels out; the calculation for the scalar diquarks and
that for the pseudoscalar mesons is essentially the same.

We can define a form factor associated with each of these two diagrams
\begin{eqnarray}
    2\,P_\mu\,F_{ab;a}(Q^2) &=&
            I_\mu^{ab;a}(P, Q) \,.
\label{eq:emff}
\end{eqnarray}
Thus the electromagnetic form factor for a diquark with flavour
structure $ab$ is given by
\begin{eqnarray}
   F_{ab}(Q^2) &=& \hat{Q_a} \; F_{ab;a} + \hat{Q_b} \; F_{ab;b} \,,
\end{eqnarray}
and the corresponding charge radius is defined as
\begin{eqnarray}
  r^2 &=& -6 \frac{F'(Q^2)}{F(Q^2)}\Bigg|_{Q^2 = 0}   \,.
\label{eq:radius}
\end{eqnarray}

For the quark propagators and diquark BSAs we use the solutions of the
rainbow DSE and ladder BSE discussed above.  Current conservation
demands that also the quark-photon vertex $\Gamma_\mu$ is dressed.  In
impulse approximation, current conservation is guaranteed as long as
the quark-photon vertex satisfies the vector Ward--Takahashi
identity~\cite{Roberts:1994hh}.  The solution of the inhomogeneous
ladder BSE
\begin{eqnarray}
\lefteqn{ \Gamma^a_\mu(p_+,p_-) \;=\; Z_2 \, \gamma_\mu - \hbox{} }
\nonumber\\ && \hbox{} \frac{4}{3}\,
        \int_q^\Lambda\! {\cal G}(k^2) \; D_{\rho\sigma}^{\rm free}(k)
    \gamma_\rho \, S(q_+)  \, \Gamma_\mu(q_+,-q_-) \, 
    S(q_-) \, \gamma_\sigma \; ,
\label{verBSE}
\end{eqnarray}
with the same effective interaction as used in the rainbow DSE for the
quark propagator does satisfy this constraint, independent of the
details of the effective interaction~\cite{Maris:emf}.

\section{\label{sec:numres}
Numerical results}
For the effective quark-antiquark interaction, we employ the
Ansatz~\cite{Maris:1999nt}
\begin{eqnarray}
\label{gvk2}
\frac{{\cal G}(k^2)}{k^2} &=&
        \frac{4\pi^2\, D \,k^2}{\omega^6} \, {\rm e}^{-k^2/\omega^2}
        + \frac{ 4\pi^2\, \gamma_m \; {\cal F}(k^2)}
        {\textstyle{\frac{1}{2}} \ln\left[\tau + 
        \left(1 + k^2/\Lambda_{\rm QCD}^2\right)^2\right]} \;.
\end{eqnarray}
with \mbox{${\cal F}(s)=(1 - \exp\frac{-s}{4 m_t^2})/s$},
\mbox{$\gamma_m=12/(33-2N_f)$}, and fixed parameters
\mbox{$m_t=0.5\,{\rm GeV}$}, \mbox{$\tau={\rm e}^2-1$},
\mbox{$N_f=4$}, and \mbox{$\Lambda_{\rm QCD} = 0.234\,{\rm GeV}$}.
The remaining parameters $\omega$ and $D$ were fitted in
Ref.~\cite{Maris:1999nt} to reproduce $f_\pi$ and the chiral
condensate; however, it turns out that the light mesons are only
sensitive to the combination $D\omega$.  The current quark masses
$m_u=m_d = 3.7~{\rm MeV}$ and $m_s = 85~{\rm MeV}$ at $\mu = 19~{\rm
GeV}$ were fitted to the pion and kaon mass respectively.  Using the
one-loop expression to scale these masses down to $\mu = 1~{\rm GeV}$
gives $m_{u,d}(1~{\rm GeV})= 5.5~{\rm MeV}$ and $m_s(1~{\rm GeV}) =
125~{\rm MeV}$.

The ultraviolet behaviour of this effective interaction is chosen to be
that of the QCD running coupling $\alpha(k^2)$; the ladder-rainbow
truncation then generates the correct perturbative QCD structure of
the set of DSEs.  In the infrared region, the interaction is
sufficiently strong to produce a realistic value for the chiral
condensate of about $(240\,{\rm GeV})^3$.

\subsection{Diquark masses}

With this model, we have solved the BSE for scalar diquarks.  In
addition to the diquark masses, we also calculated the axial-vector
projection of the diquark wave function at the origin, which is the
analogue of the leptonic decay constant of the mesons.  To be
specific, we define an axial-vector colour and flavour anti-triplet
operator, and calculate
\begin{eqnarray}
  f_{D,ab} &=& Z_2\, \int_q^\Lambda\! {\rm Tr}\bigg[ 
    S_a(q_+) \, \Big(\Gamma_{D,ab}(q_+,-q_-) \, C\Big) \, S_b(q_-) \, 
    \gamma_5 \gamma_\mu \bigg] \;.
\end{eqnarray} 
This quantity $f_{D,ab}$ is a purely theoretical object, and does not
correspond to any physical decay constant of the diquarks.  However,
it is renormalisation-point-independent, just like the meson decay
constants.  The difference with the expression for the meson decay
constants is the absence of a factor of $N_c$.  Hence, with this
definition, we expect $f_{D,ab}$ to be about $\sqrt{3}$ times smaller
than the corresponding pion and kaon decay constants (remember that
this factor of $N_c$ is also absent from the diquark normalisation
condition).  Any deviations from this expectation would signal
dynamical differences between the relative momentum behaviour of the
diquark and meson BSAs.  Our results are given in
Table~\ref{tab:mass}.
\begin{table}[bth]
\beforetab
\begin{tabular}{llccccccccl}
\firsthline
\multicolumn{2}{l}{ input parameters} & \multicolumn{5}{c}{ meson }
    & \multicolumn{4}{c}{ diquark }  \\ 
\midhline
$\omega$ & $D$ $[{\rm GeV}^2]$ & $m_\pi$ & $f_\pi$ & $m_K$ & $f_K$ 
        & & $m_{ud}$ & $\sqrt{3}f_{ud}$ & $m_{qs}$ & $\sqrt{3}f_{qs}$ \\ 
\midhline
0.30 & 1.25  & 0.138 & 0.1305 & 0.494 & 0.1537 & & 0.988 & 0.124 & ---  & ---    \\
0.40 & 0.93  & 0.138 & 0.1305 & 0.495 & 0.1537 & & 0.821 & 0.128 & 1.10 & 0.17(2)\\
0.42 & 0.888 & 0.138 & 0.1308 & 0.496 & 0.1541 & & 0.791 & 0.129 & 1.07 & 0.170  \\
0.45 & 0.829 & 0.139 & 0.1306 & 0.496 & 0.1544 & & 0.744 & 0.128 & 1.02 & 0.163  \\
0.50 & 0.79  & 0.138 & 0.1307 & 0.495 & 0.1558 & & 0.688 & 0.127 & 0.96 & 0.158  \\
\lasthline
\end{tabular}
\aftertab \captionaftertab[]{\label{tab:mass} 
Overview of our results for the pseudoscalar meson and scalar diquark
masses and decay constants for different model parameters.  All
quantities except $D$ are in GeV.  Numerical errors are of the order
of 1\% for the mesons, and 2\% to 5\% for the diquarks (unless noted
explicitly), with the heavier states having the larger numerical
uncertainty.}
\end{table}

The light meson properties seem to be rather independent of the
details of the effective interaction, as long as the interaction
generates the observed amount of chiral symmetry
breaking~\cite{Maris:1999nt}.  This can be achieved by keeping
$D\omega \approx (0.72~{\rm GeV})^3$ fixed: the light meson masses
(and other calculated properties) remain unchanged for $0.3~{\rm GeV}
< \omega < 0.5~{\rm GeV}$, provided $D$ is adjusted accordingly, see
Table~\ref{tab:mass}.  However, the diquark masses are more sensitive
to details of the effective interaction: a 10\% increase in $\omega$
while keeping $D\omega$ constant reduces the scalar diquark mass by
about 10\%.  Furthermore, if $\omega$ becomes smaller than about
0.38~GeV (and $D$ larger than about 1~GeV$^2$), we can no longer find
a stable $us$ or $ds$ diquark.

Also some of the heavier mesons are sensitive to the range of the
strong infrared attraction in the model.  However, this dependence on
$\omega$ is not uniform.  For example, changing $\omega$ from 0.4~GeV
to 0.5~GeV reduces the mass of the scalar meson by
10\%~\cite{Maris:2002yu}.  In contrast, the mass of the first radial
excitation of the pion is increased by 14\% under a similar
change~\cite{Holl:2004fr}.  The vector mesons are insensitive
to these changes in $\omega$ and $D$~\cite{Maris:1999nt}.

A complication with these calculations is that the rainbow truncation
of the quark DSE in general leads to a pair of complex conjugate
singularities in the quark propagator~\cite{sing,Alkofer:2003jj}.
These singularities are most likely artefacts of the model, and
progress is being made in eliminating these
artefacts~\cite{Alkofer:2003jj,andreas}.  Within the model under
consideration here, the exact location of these singularities in the
complex plane depends on the parameters~\cite{andreas}, but typically
the integration domain spanned by the BSE encroaches on these
singularities for bound state masses of slightly above 1~GeV.  This
limits the exploration of heavier states, and it also limits the range
of $Q^2$ for which we can reliably calculate the electromagnetic form
factor.

\subsection{Electromagnetic form factors}

Next, we use the obtained diquark BSAs to calculate the
electromagnetic form factors.  For completeness, we also include the
charge radii for the pseudoscalar mesons, illustrating that the pion
and kaon properties are indeed insensitive to the details of the
interaction.  For the various 
mesons we have
\begin{eqnarray}
  F_{\pi}(Q^2) &=& 
\frac{2}{3}F_{u\bar{d};u}(Q^2) + \frac{1}{3} F_{u\bar d;\bar d}(Q^2) \,,
\\
  F_{K^+}(Q^2) &=& 
\frac{2}{3} F_{u\bar{s};u}(Q^2) + \frac{1}{3} F_{u\bar s;\bar s}(Q^2) \,,
\\
  F_{K^0}(Q^2) &=& 
-\frac{1}{3} F_{d\bar{s};d}(Q^2) + \frac{1}{3} F_{d\bar s;\bar s}(Q^2) \,,
\end{eqnarray}
whereas for the diquarks we have
\begin{eqnarray}
  F_{ud}(Q^2) &=& 
\frac{2}{3} F_{ud;u}(Q^2) - \frac{1}{3} F_{ud;d}(Q^2) \,,
\\
  F_{us}(Q^2) &=& 
\frac{2}{3} F_{us;u}(Q^2) - \frac{1}{3} F_{us;s}(Q^2) \,,
\\
  F_{ds}(Q^2) &=& 
-\frac{1}{3} F_{ds;d}(Q^2) - \frac{1}{3} F_{ds;s}(Q^2) \,.
\end{eqnarray}
Impulse approximation, in combination with the rainbow-ladder
truncation, ensures current conservation: each of the form factors
$F_{q\bar{q};q}(Q^2)$ and $F_{qq;q}(Q^2)$ is one at $Q^2 = 0$,
independent of the flavour label $q$.  Thus we obtain the correct
(integer) charges for the physical mesons, whereas for the diquarks we
get fractional charges $1/3$ for the $ud$ and $us$ diquark, and $-2/3$
for the $ds$ diquark.  Note however that these diquark states are
coloured objects, and therefore confined, and these fractional charges
cannot be observed directly.  With the above definition of the
``physical'' form factors, $F_{ab}(0)$ gives the charge of the state
under consideration.  This (fractional) charge is divided out in the
definition of the charge radius, see Eq.~(\ref{eq:radius}).

\begin{table}[bth]
\beforetab
\begin{tabular}{llccccccccccc}
\firsthline
\multicolumn{2}{l}{ input } & \multicolumn{6}{c}{ meson }
    & \multicolumn{5}{c}{ diquark }  \\ 
\midhline
$\omega$ & $D$
& $r^2_\pi$    & $r^2_{q\bar{s};q}$ & $r^2_{q\bar{s};\bar{s}}$ 
               & $r^2_{K^+}$ & $r^2_{K^0}$ 
& & $r^2_{ud}$& $r^2_{qs;q}$ & $r^2_{qs;s}$ 
               & $r^2_{us}$ & $r^2_{ds}$ \\ 
\midhline
0.40 & 0.93  & 0.433 & 0.460 & 0.209 & 0.376 & -.084 &
          & 0.51 & ---  & ---  & --- & --- \\
0.42 & 0.888 & 0.433 & 0.458 & 0.210 & 0.375 & -.083 &
          & 0.51 & 0.45 & 0.27 & 0.63 & 0.36\\
0.45 & 0.829 & 0.438 & 0.460 & 0.211 & 0.377 & -.083 &
          & 0.51 & 0.49 & 0.27 & 0.71 & 0.38 \\
0.50 & 0.79  & 0.441 & 0.455 & 0.211 & 0.374 & -.081 &
          & 0.51 & 0.49 & 0.27 & 0.71 & 0.38 \\
\lasthline
\end{tabular}
\aftertab \captionaftertab[]{\label{tab:radii} 
Overview of our results for the pseudoscalar meson and scalar diquark
charge radii squared, all in ${\rm fm}^2$ (the flavour label $q$
stands for both $u$ and $d$ quarks).  The estimate of the combined
numerical errors is less than $0.01~{\rm fm}^2$ for the meson charge
radii, about $0.01~{\rm fm}^2$ for the $ud$ radii, and about 5\% to
10\% for the $us$ and $ds$ radii.  For $\omega= 0.4~{\rm GeV}$, $D =
0.93~{\rm GeV}^2$, numerical difficulties due to the complex
singularities in the quark propagator prevent an accurate
determination of the charge radius for the $us$ and $ds$ diquarks.}
\end{table}
Our numerical results for the charge radii are given in
Table~\ref{tab:radii}.  We work in the isospin symmetry limit, and
thus we do not discriminate between the $u$ and the $d$ quark, except
for the obvious difference in their electromagnetic charge.  As
already mentioned, the pion and kaon charge radii are insensitive to
variations of the parameters as long as the integrated infrared
strength is kept fixed.  Also the diquark charge radii turn out to be
relatively insensitive to these variations, despite the fact that the
diquark masses change considerably over this range in the parameters.
This is at least partly due to the fact that the vector meson masses
are also insensitive to these variations in the parameters, and they
show up as poles in the dressed quark-photon vertex.

\begin{figure}[tbh]
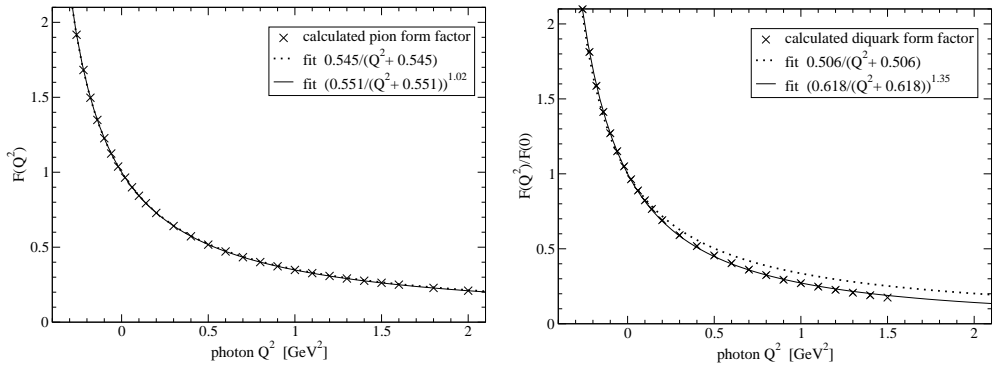

\begin{center}
\parbox{66mm}{
\epsfig{file=emfpioncorrected.eps,width=64mm}}
\parbox{66mm}{
\epsfig{file=emfuucorrected.eps,width=64mm}}
\end{center}
\caption[]{\label{fig:emfresults}
The pion (left) and $ud$ diquark (right) form factor, together with
the fitted forms, for $\omega=0.42~{\rm GeV}$, $D = 0.888~{\rm
GeV}^2$.  The analytic forms are fitted in the range $-0.3~{\rm GeV}^2
< Q^2 < 1.5~{\rm GeV}^2$. }
\end{figure}
Indeed, all of the form factors do show a pole-like structure in the
timelike region, see Fig.~\ref{fig:emfresults}.  For the pion, a
monopole fit to the form factor
\begin{eqnarray}
  F(Q^2) &=& \frac{M^2}{Q^2 + M^2}
\end{eqnarray}
using 27 calculated values of $F(Q^2)$ with $Q^2$ in the range
$-0.4~{\rm GeV}^2 < Q^2 < 1.6~{\rm GeV}^2$ gives a mass-scale that is
very close to the $\rho$ meson mass in this model, $m^2_\rho =
0.55~{\rm GeV}^2$, with a $\chi^2$ of about 0.001.  Again, these
results are almost independent of the details of the interaction, as
can be seen from Table~\ref{tab:monopoles}.

For mesons (and diquarks) involving one strange quark and one up or
down quark, one has to choose whether to attempt a monopole fit to the
form factors corresponding to the individual diagrams of
Fig.~\ref{fig:triangle}, or to the physical form factors.  The neutral
kaon form factor cannot be fitted by a simple monopole; however, it
might be fit by a difference of two monopoles.  Indeed, the vector
meson dominance [VMD] mechanism suggests that one should fit the
individual diagrams of Fig.~\ref{fig:triangle}: it is the vector meson
pole in the quark-photon vertex that is at the origin of the success
of VMD models.  Thus, for the kaon, we have fitted
$F_{u\bar{s};u}(Q^2)$ and $F_{u\bar s ; \bar s}(Q^2)$ over the same
range, and the agreement with a monopole is reasonable, though not as
good as for the pion; the $\chi^2$ value is about 0.01.  Furthermore,
the masses deviate somewhat more from the vector mesons $m^2_\rho =
0.55~{\rm GeV}^2$ and $m^2_\phi = 1.15$~GeV$^2$, see
Table~\ref{tab:monopoles}.  As the pole positions are approached,
these form factors become increasingly better represented as VMD
monopoles with the $\rho$ and $\phi$
masses~\cite{Maris:emf,Jarecke:2002xd}.
\begin{table}[bth]
\beforetab
\begin{tabular}{llcccccccccc}
\firsthline
\multicolumn{2}{l}{ input parameters} & \multicolumn{3}{c}{ meson }
    & \multicolumn{4}{c}{ diquark }  
    & \multicolumn{3}{c}{ $ud$ diquark best fit }  \\
\midhline
$\omega$ $[{\rm GeV}]$ & $D$ $[{\rm GeV}^2]$ 
      & pion & $q\bar{s};q$ & $q\bar{s};q$
      & & $ud$ & $qs;q$ & $qs;s$ 
      & & mass & power   \\
\midhline
0.40 & 0.93  & 0.545 & 0.529 & 1.116 & & 0.51 & ---  & ---  & & 0.62 & 1.35 \\
0.42 & 0.888 & 0.545 & 0.530 & 1.109 & & 0.51 & 0.4  & 0.8  & & 0.62 & 1.35 \\
0.45 & 0.829 & 0.540 & 0.527 & 1.100 & & 0.50 & 0.54 & 0.85 & & 0.63 & 1.38 \\
0.50 & 0.79  & 0.538 & 0.530 & 1.103 & & 0.50 & 0.51 & 0.89 & & 0.62 & 1.37 \\
\lasthline
\end{tabular}
\aftertab \captionaftertab[]{\label{tab:monopoles} 
Estimates for monopole fit masses $M^2$ in GeV$^2$
(the flavour label $q$ stands for both $u$ and $d$ quarks).}
\end{table}

For the diquarks, a monopole fit is not very good, as can be seen from
the right panel of Fig.~\ref{fig:emfresults}; the $\chi^2$ of the
monopole fit is around 0.1 for the $ud$ diquark.  A much better fit
can be achieved by using the two-parameter fit form
\begin{eqnarray}
  F(Q^2) &=& \left(\frac{M^2}{Q^2 + M^2}\right)^a
\end{eqnarray}
which gives a $\chi^2$ of 0.001.  The preferred power appears to be
$a=1.4$, independent of the details of the interaction.  However, this
does depend somewhat on the $Q^2$ range that is used in the fitting.
Close to the $\rho$ pole in the timelike region, it does approach a
pure VMD monopole form.  A similar fit to the calculated pion form
factor gives a power of $1.02$, which is hardly an improvement over
the simple monopole fit.  These fits with an arbitrary power can only
be good for low and intermediate values of $Q^2$, since we know that
asymptotically, the pseudoscalar meson and scalar diquark form factors
vanish like $1/Q^2$.

For the $qs;q$ and $qs;s$ form factors, with $q$ standing for $u$ or
$d$ quarks, the monopole masses in Table~\ref{tab:monopoles} are an
estimate based on the low-$Q^2$ behaviour only, rather than a fit over
the same range of $Q^2$.  Complications with the analytic continuation
of the quark propagator solution necessary for these calculations, and
in particular the complex singularities in the quark propagator,
prevent us from obtaining the form factor for these heavier states
away from $Q^2 \approx 0$.

In conclusion, we can say that simple VMD models work very well for
the pion form factor, but they are less and less accurate for the
heavier mesons.  Nevertheless, the monopole forms for the diquarks are
a reasonable approximation to these microscopic form factor
calculations, and might be useful for applications in baryon
calculations.  An interesting consequence of these monopole fits is
that the $us$ diquark form factor could potentially have a zero
crossing in the spacelike region: if the mass-scale $M^2$ in the
$F_{us;s}$ fit is more than twice as large as the mass scale $M^2$ in
the $F_{us;u}$ fit, the ``physical'' form factor $\frac{2}{3}
F_{us;u}(Q^2) - \frac{1}{3} F_{us;s}(Q^2)$ will approach zero from
below for $Q^2 \to \infty$.  (In terms of VMD masses that means if
$M_\phi^2 > 2 M_\rho^2$.)  However, both VMD and our current
calculations indicate that this is not the case for realistic up,
down, and strange quark masses.

\section{\label{sec:conc}
Concluding remarks}
The calculated charge radius of the $ud$ scalar diquark, $r_{ud} =
0.71~{\rm fm}$, is 8\% larger than the pion charge radius, $r_\pi =
0.66~{\rm fm}$.  This suggests that these diquarks are about 8\%
larger in size than pions, which agrees with the observation that the
diquark BSA is about the same amount narrower in momentum space than
the pion BSA~\cite{Maris:2002yu}.  These results appear to be
insensitive to the details of the effective quark-quark interaction
kernel, even though the actual diquark masses are sensitive to details
of the interaction.  Our results disagree with calculations within the
Nambu--Jona-Lasinio [NJL] model~\cite{Weiss:1993kv}, which suggested
that the electromagnetic form factors of the pion and $ud$ scalar
diquark are almost identical.  Furthermore, the obtained charge radius
in those NJL calculations is significantly smaller than the measured
pion charge radius.

Our result for the $ud$ scalar diquark charge radius is about 20\%
smaller than the experimental proton charge radius,
$r_{\hbox{\scriptsize proton}} = 0.87~{\rm fm}$.  If we take these
radii as an indication of their physical size, it is indeed quite
plausible to picture baryons as bound states of a quark and a diquark.
For quantitatively reliable calculations of electromagnetic properties
of nucleons however, we need not only the scalar diquark, but also the
axial-vector diquark properties.

Recently, it has been suggested that pentaquarks might be composed of
two diquarks (coupled together in a colour triplet) coupled to an
antiquark~\cite{Jaffe:2003sg}.  Another possibility involving diquarks
would be a (colour anti-triplet) diquark coupled to a triquark
($qq\bar{q}$) in a colour triplet state~\cite{Karliner:2003dt}.  One
would expect pentaquarks to be somewhat larger in size than the
ordinary baryons.  Based on the charge radius of the diquarks and that
of the proton, we would estimate the radius of a pentaquark (in a
diquark-diquark-antiquark configuration) to be $r_{\hbox{\scriptsize
pentaquark}} \approx 1~{\rm fm}$, or about 15\% larger than that of
the proton.

In the timelike region, all form factors can be approximated quite
well by a VMD like monopole.  For increasing spacelike values of
$Q^2$, a VMD monopole becomes less and less accurate, in particular
for heavier mesons and diquarks.  For moderate spacelike values of
$Q^2$, the $ud$ diquark form factor may be interpolated by $[M^2/(Q^2
+ M^2)]^{1.4}$, with a mass of about $M \approx 0.79~{\rm GeV}$.
Asymptotically however, we expect this form factor to vanish like
$1/Q^2$, just like the pion form factor.

\begin{acknowledge}
I would like to thank Arne Hoell, Andreas Krassnigg, Craig Roberts,
Eric Swanson, and Peter Tandy for useful discussions.  Part of the
computations were performed on the National Science Foundation
Terascale Computing System at the Pittsburgh Supercomputing Center.
\end{acknowledge}


\end{document}